\documentclass[onecolumn,showpacs,floatfix,11pt,nofootinbib]{revtex4}
\usepackage[dvips]{graphics,color}
\usepackage{epsfig}\usepackage{float}
\RequirePackage{amssymb}
\usepackage{makeidx}
\usepackage[brazil]{babel}
\usepackage[latin1]{inputenc}
\usepackage{graphicx}
\usepackage{indentfirst}
\usepackage{color}
\newcommand{\be}{\begin{equation}}
\newcommand{\ee}{\end{equation}}

\newcommand{\ba}{\begin{eqnarray}}
\newcommand{\ea}{\end{eqnarray}}

\begin{document}

\title{On the Consistence Conditions to Braneworlds Sum Rules within Scalar-Tensor Gravity for Arbitrary Dimensions}

\author{M. C. B. Abdalla$^{1}$}
\email{mabdalla@ift.unesp.br}
\author{M. E. X. Guimarães$^{2}$}
\email{emilia@if.uff.br}
\author{J. M. Hoff da Silva$^{1}$}
\email{hoff@ift.unesp.br}

\affiliation{1. Instituto de F\'{\i}sica Te\'orica, Universidade
Estadual Paulista, Rua Pamplona 145 01405-900 S\~ao Paulo, SP,
Brazil}

\affiliation{2. Instituto de F\'{\i}sica, Universidade Federal
Fluminense, Av. Gal. Milton Tavares de Souza s/n, 24210-346,
Niter\'oi-RJ, Brazil}


\begin{abstract}
We derive an one-parameter family of consistence conditions to
braneworlds in the Brans-Dicke gravity. The sum rules are
constructed in a completely general frame and they reproduce the
conditions already obtained in General Relativity theory just by
using a right limit of the Brans-Dicke parameter.
\end{abstract}
\maketitle

\section{Introduction}

In the recent years braneworld models are consolidating a new
branch of high energy physics. Among many interesting physical
insights \cite{Maartens}, they provide an elegant solution to the
hierarchy problem \cite{muitos,RSI,RSII}. The development of new
braneworld models is increasing in direct proportion to the
possibilities raised in the scope of a extra dimensional world.

On the other hand, formal advances of string theory point into a
scalar tensorial theory as an effective theory which gives the
right approach to the gravitational phenomena \cite{novos}.
Unification models such as supergravity, superstrings and M-theory
\cite{GSW} effectively predict the existence of a scalar
gravitational field acting as a mediator of the gravitational
interaction together with the usual rank-2 tensorial field. In
this context, the analysis of braneworld consistence conditions
are indeed necessary. Models with large extra dimensions, in which
our universe is usually performed as a brane, have typical size
scales much larger than the scales involved in quantum gravity.
Therefore, the sum rules arise manly in a gravitational theory
scenario. These sum rules were obtained in ref. \cite{GKL} for
five dimensions in General Relativity\footnote{It is  directly
applied in the Randall-Sundrum model \cite{RSI,RSII}.} (GR) and
extended to an arbitrary number of dimensions in ref.
\cite{Leblond}, also in GR.

The main purpose of this work is to generalize the braneworld sum
rules, already obtained in the GR set up, to the case of the
scalar tensorial gravity. In particular, we work with the
Brans-Dicke (BD) theory \cite{BD}, since it is the simplest
scalar-tensor theory we have in the literature. The paper is
structured as follows: in the next Section we establish the
notation and then, generalize to the case of the BD gravity
framework. After arriving to the main general results we
particularize the analysis to braneworld models obtained in ref.
\cite{AG}. In the final Section we conclude with some remarks.

\section{Generalized Sum Rules}

In this section we obtain the generalized braneworld sum rules
within the BD gravity framework. First we fix the notation and,
since the geometrical part of the Einstein's equation is the same
as in the BD case, we derive the basic setup to obtain the
consistence conditions following the prescription  of ref.
\cite{Leblond}. After some general considerations, we extend the
analysis in order to incorporate the scalar field terms
(generically called as dilaton here). Basically, we obtain an
one-parameter family of consistence conditions to the BD case by a
circular integration in the internal compact space.

\subsection{Notation and Conventions}

Following the standard notation used in refs. \cite{GKL,Leblond},
we analyze a D-dimensional bulk spacetime endowed with a
non-factorable geometry, which metric is given by \be
ds^{2}=G_{AB}dX^{A}dX^{B}=W^{2}(r)g_{\alpha\beta}dx^{\alpha}dx^{\beta}+g_{ab}(r)dr^{a}dr^{b}\label{1},
\ee where $W^{2}(r)$ is the warp factor, assumed to be a smooth
integrable function, $X^{A}$ denotes the coordinates of the full
D-dimensional spacetime, $x^{\alpha}$ stands for the $(p+1)$
non-compact coordinates of the spacetime and $r^{a}$ labels the
$(D-p-1)$ directions in the internal compact space\footnote{As an
example, if $D=5$, $p=3$ and $W(r)=e^{-2k|r|}$ one arrives to the
Randall-Sundrum model.}. Note that this type of metric encodes the
possibility of existing q-branes $(q>p)$ \cite{Leblond}. In this
case, the $(q-p)$ extra dimensions are compactified on the brane
and constitute part of the internal space. This possibility is
important in the hybrid compactification models context.

The D-dimensional spacetime Ricci tensor can be related with the
brane Ricci tensor as well as with the internal space partner by
the equations \cite{GKL} \be
R_{\mu\nu}=\bar{R}_{\mu\nu}-\frac{g_{\mu\nu}}{(p+1)W^{p-1}}\nabla^{2}W^{p+1},\label{2}
\ee and \be
R_{ab}=\tilde{R}_{ab}-\frac{p+1}{W}\nabla_{a}\nabla_{b}W,\label{3}\ee
where $\tilde{R}_{ab}$, $\nabla_{a}$ and $\nabla^{2}$ are
respectively the Ricci tensor, the covariant derivative and the
Laplacian operator constructed by the internal space metric
$g_{ab}$. $\bar{R}_{\mu\nu}$ is the Ricci tensor derived from
$g_{\mu\nu}$. Let us denote the three curvature scalars by
$R=G^{AB}R_{AB}$, $\bar{R}=g^{\mu\nu}\bar{R}_{\mu\nu}$ and
$\tilde{R}=g^{ab}\tilde{R}_{ab}$. Therefore, the traces of
equations (\ref{2}) and (\ref{3}) give \be
\frac{1}{p+1}\Big(W^{-2}\bar{R}-R^{\mu}_{\mu}\Big)=pW^{-2}\nabla
W\cdot \nabla W+W^{-1}\nabla^{2}W \label{4}\ee and \be
\frac{1}{p+1}\Big(\tilde{R}-R_{a}^{a}\Big)=W^{-1}\nabla^{2}W,\label{5}
\ee where $R^{\mu}_{\mu}\equiv W^{-2}g^{\mu\nu}R_{\mu\nu}$ and
$R^{a}_{a}\equiv g^{ab}R_{ab}$ (in such a way that
$R=R^{\mu}_{\mu}+R^{a}_{a}$). It is not difficult to see that, if
$\xi$ is an arbitrary constant, \be \nabla \cdot (W^{\xi}\nabla
W)=W^{\xi+1}(\xi W^{-2}\nabla W\cdot \nabla W+W^{-1}\nabla^{2}
W)\label{6}.\ee The combination of the equations (\ref{4}),
(\ref{5}) and (\ref{6}) leads to \be \nabla \cdot (W^{\xi}\nabla
W)=\frac{W^{\xi+1}}{p(p+1)}[\xi\big(W^{-2}\bar{R}-R^{\mu}_{\mu}\big)+(p-\xi)\big(\tilde{R}-R^{a}_{a}\big)].\label{7}
\ee This is a very important relation. In particular, it will
provide the consistence conditions since the left-hand side must
vanish under a closed integration over the internal compact space.

\subsection{The Scalar-Tensor Gravity Case}

Once established the usual notation and conventions, it is time to
look at the dilaton field $(\phi)$ contributions. The
Einstein-Brans-Dicke (EBD) equation is given by
\begin{eqnarray}
R_{MN}-\frac{1}{2}G_{MN}R&=&\left.
\frac{8\pi}{\phi}T_{MN}+\frac{w}{\phi^{2}}\Big(\nabla_{M}\phi\nabla_{N}\phi-
\frac{1}{2}\nabla_{A}\phi\nabla^{A}\phi G_{MN}
\Big)\right.\nonumber\\&+&\left.\frac{1}{\phi}\Big(\nabla_{M}\nabla_{N}\phi-\frac{8\pi}{3+2w}TG_{MN}
\Big)\right.,\label{8}
\end{eqnarray}
where $T_{MN}$ is the matter stress-tensor (everything except
$\phi$), $T$ is the trace and $w$ the BD parameter. We remark that
the scalar part of the BD set of equations was already taken into
account in the last term of the right-hand side of eq. (\ref{8}).
From eq. (\ref{8}) it is easy to note that \be
R_{MN}=\frac{8\pi}{\phi}\Big(T_{MN}+\frac{2(1+w)T}{(2-D)(3+2w)}G_{MN}\Big)+\frac{w}{\phi^{2}}\nabla_{M}\phi
\nabla_{N}\phi+\frac{1}{\phi}\nabla_{M}\nabla_{N}\phi.\label{9}
\ee Now, if we call $T^{\mu}_{\mu}\equiv
W^{-2}g^{\mu\nu}T_{\mu\nu}$ $(T=T^{\mu}_{\mu}+T^{m}_{m})$, it is
possible to express $R^{\mu}_{\mu}$ and $R^{m}_{m}$ by
\begin{eqnarray}
R^{\mu}_{\mu}&=&\left.\frac{8\pi}{\phi(D-2)(3+2w)}\Big((3D+2w(D-p-3)-2(p+1))T^{\mu}_{\mu}-2(1+w)(p+1)T^{m}_{m}\Big)
\right.\nonumber\\&+&\left.\frac{wW^{-2}}{\phi^{2}}\nabla^{\nu}\phi\nabla_{\nu}\phi+\frac{W^{-2}}{\phi}\nabla^{\nu}\nabla_{\nu}\phi,\right.
\label{10}\end{eqnarray} and
\begin{eqnarray}
R^{m}_{m}&=&\left.\frac{8\pi}{\phi(D-2)(3+2w)}\Big((D+2w(p-1)-2(p-2))T^{m}_{m}-2(1+w)(D-p-1)T^{\mu}_{\mu}\Big)
\right.\nonumber\\&+&\left.\frac{w}{\phi^{2}}\nabla^{m}\phi\nabla_{m}\phi+\frac{1}{\phi}\nabla^{m}\nabla_{m}\phi.\right.
\label{11}\end{eqnarray} Inserting these last two
equations\footnote{It is important to remark that when one takes
the limit $w\rightarrow \infty$ $(\phi\rightarrow 1/G_{N})$ the
expressions $R^{\mu}_{\mu}$ and $R^{m}_{m}$ recover the case
analyzed in General Relativity.} in (\ref{7}) one has
\begin{eqnarray}
\nabla \cdot (W^{\xi}\nabla W)&=&\left.
\frac{W^{\xi+1}}{p(p+1)}\Bigg(\xi
W^{-2}\bar{R}+(p-\xi)\tilde{R}-\frac{8\pi}{\phi(D-2)(3+2w)}\right.\nonumber\\&\times&
\left.\Bigg(T^{\mu}_{\mu}[\xi(5D+4w(D-p-2)-2(2p+5))-2p(1+w)(D-p-1)]\right.\nonumber\\
&+&\left.T^{m}_{m}[\xi(-4wp-D+2(1-2p))+p(D-2(p-2)+2w(p-1))]\Bigg)\right.\nonumber\\
&-&\left.\frac{w}{\phi^{2}}[\xi
W^{-2}\nabla^{\mu}\phi\nabla_{\mu}\phi+(p-\xi)\nabla^{m}\phi\nabla_{m}\phi]-\frac{1}{\phi}[\xi
W^{-2}\nabla^{\mu}\nabla_{\mu}\phi\right.\nonumber\\
&+&\left.(p-\xi)\nabla^{m}\nabla_{m}\phi]\Bigg).\label{12}\right.
\end{eqnarray}

An important characteristic about braneworld models in the BD
theory is that, generally, the scalar field depends only on the
large extra dimensions \cite{AG}. This type of dependence is
indeed useful since the projected EBD equation on the brane leads
to important subtle modifications but still resembles the
Einstein's equation \cite{AG2}. Keeping it in mind, let us
considerer hereon the $\nabla_{\mu}\phi=0$ case. In an internal
compact space the follow identity is respected \be \oint \nabla
\cdot (W^{\xi}\nabla W)=0,\label{13}\ee therefore from eq.
(\ref{12}) we have
\begin{eqnarray}&&
\left.\oint
\frac{W^{\xi+1}}{\phi}\Bigg(T^{\mu}_{\mu}[\xi(5D+4w(D-p-2)-2(2p+5))-2p(1+w)(D-p-1)]\right.\nonumber\\&+&\left.
T^{m}_{m}[\xi(-4wp-D+2(1-2p))+p(D-2(p-2)+2w(p-1))]-\Big(\frac{8\pi}{\phi}\Big)^{-1}(D-2)(3+2w)\right.\nonumber\\&\times&\left.
[\xi
W^{-2}\bar{R}+(p-\xi)\tilde{R}]+\Big(\frac{8\pi}{\phi}\Big)^{-1}(D-2)(3+2w)\Big(\frac{w}{\phi^{2}}\nabla^{m}\phi\nabla_{m}\phi
+\frac{1}{\phi}\nabla^{m}\nabla_{m}\phi\Big)\Bigg)=0\label{14}.\right.
\end{eqnarray}
The above equation provides an one-parameter family of consistence
conditions in arbitrary dimensions in the scope of the BD gravity.
It is important to stress that if one reintroduces the
$(3+2w)^{-1}$ factored term and takes $w\rightarrow \infty$
$(\phi\rightarrow 1/G_{N})$, by the usual L'Hôpital limit, the
case analyzed in General Relativity is recovered as
expected\footnote{See, for instance, eq. (13) of reference
\cite{Leblond}.}.

Equation (\ref{14}) is quite general and self-consistent. However,
in that form it is not very useful. Going forward, we rewrite eq.
(\ref{14}) in terms of an adequate energy-momentum tensor form,
after what the generalized sum rules may be applied to any
particular model. We shall use the same stress-tensor {\it Ansatz}
of ref. \cite{Leblond}, since it is very complete. So, we write
the stress-tensor in the form \be T_{MN}=-\Lambda
G_{MN}-\sum_{i}T_{q}^{(i)}P[G_{MN}]_{q}^{(i)}\Delta^{(D-q-1)}(r-r_{i})+\tau_{MN},\label{15}
\ee where $\Lambda$ is the cosmological constant, $T_{q}^{(i)}$ is
the $i^{th}$ q-brane tension, $\Delta^{(D-q-1)}(r-r_{i})$ is the
covariant combination of delta functions which positions the
brane\footnote{For a complete discussion about the expression of
$\Delta^{(D-q-1)}(r-r_{i})$ see the Appendix of ref.
\cite{Leblond}.}, $P[G_{MN}]_{q}^{(i)}$ is the pull-back of the
bulk metric and any other matter contribution is due $\tau_{MN}$.
From this {\it Ansatz} one obtains \be
T^{\mu}_{\mu}=-(p+1)\Lambda+\tau^{\mu}_{\mu}-\sum_{i}T_{q}^{i}
\Delta^{(D-q-1)}(r-r_{i})(p+1),\label{16} \ee and \be
T^{\mu}_{\mu}=-(D-p-1)\Lambda+\tau^{m}_{m}-\sum_{i}T_{q}^{i}
\Delta^{(D-q-1)}(r-r_{i})(q-p)\label{17}.\ee Now, substituting
these expressions in equation (\ref{14}) we get, after some
algebra, the following form to the consistence conditions
\begin{eqnarray}&&
\left.\oint
\frac{W^{\xi+1}}{\phi}\Bigg(-\Lambda(c(p+1)+bD)-\sum_{i}(cp+a+bq)T_{q}^{(i)}
\Delta^{(D-q-1)}(r-r_{i})+aT^{\mu}_{\mu}+bT^{m}_{m}\right.\nonumber\\&-&\left.\Big(\frac{8\pi}{\phi}\Big)^{-1}(D-2)(3+2w)
[\xi
W^{-2}\bar{R}+(p-\xi)\tilde{R}]+\Big(\frac{8\pi}{\phi}\Big)^{-1}(D-2)(3+2w)\right.\nonumber\\&\times&\left.
\Big(\frac{w}{\phi^{2}}\nabla^{m}\phi\nabla_{m}\phi
+\frac{1}{\phi}\nabla^{m}\nabla_{m}\phi\Big)\Bigg)=0,\right.\label{18}
\end{eqnarray}
where $a\equiv \xi[5D+4w(D-p-2)-2(2p+5)]-2p(1+w)(D-p-1)$, $b\equiv
\xi[-4wp-D+2(1-2p)]+p[D-2(p-2)+2w(p-1)]$ and $c\equiv a-b$.

The above considerations can be considerably simplified if the
scalar field has a power-law form. Note that if \be
\phi=[(w+1)(C_{1}r+C_{2})]^{\frac{1}{w+1}},\label{19}\ee where
$C_{1}$ and $C_{2}$ are integration constants, the last term of
the right-hand side of equation (\ref{18}) disappears. Therefore,
there exist a particular solution of the dilaton field in which
the ``kinetic'' term does not intervene in any consistence
condition. In other words, in this solution the derivative part of
the scalar field does not impose an additional constraint to the
model. It is a quite interesting characteristic. However, it is
not an usual solution to the scalar field \cite{AG}. Moreover, in
models within BD gravity the explicit form of the $\phi$ field
composes the warp factor. So, solutions as (\ref{19}) are not
potentially interesting to the solution of the hierarchic problem.
Hereon, we shall still consider the last term of eq. (\ref{18}).

\subsection{A Particular Case Example}

In order to study a specific example in BD theory, let us
particularize our analysis to braneworld models inspired in one of
those previously found in this framework \cite{AG}. In the models
proposed in \cite{AG}, all the bulk-brane structure were generated
by using local and global cosmic string as sources. The final
scenario is composed by 4-branes embedded in a bulk of
six-dimensions. Therefore, they are models of hybrid
compactification. The on-brane dimension is compactified into a
$S^{1}$ cycle and the dilaton field depends only on the large
transverse dimension. With these specifications, we have $D=6$,
$p=3$ and $q=4$, and consequently, $a=2\xi(15+2w)-2(17+6w)$,
$b=-4\xi(4+3w)+12(1+w)$ and $c=2\xi(23+8w)-2(23+12w)$. Then, the
equation (\ref{18}) gives
\begin{eqnarray}&&
\left. \oint \frac{W^{\xi+1}}{\phi}\Bigg(-4\Lambda
[35\xi-w(\xi+3)-14]-2\sum_{i}[26\xi+w(\xi-9)-31]T_{4}^{(i)}\Delta^{(1)}(r-r_{i})\right.\nonumber\\&+&
\left.
\tau^{\mu}_{\mu}[\xi(15+2w)-(17+6w)]+\tau^{m}_{m}[-2\xi(4+3w)+6(1+w)]-2(3+2w)\Big(\frac{8\pi}{\phi}\Big)^{-1}
\right.\nonumber\\&\times&\left. [\xi
W^{-2}\bar{R}+(3-\xi)\tilde{R}]+2(3+2w)(3-\xi)\Big(\frac{8\pi}{\phi}\Big)^{-1}\Big(\frac{w}{\phi^{2}}\nabla^{m}\phi\nabla_{m}\phi
+\frac{1}{\phi}\nabla^{m}\nabla_{m}\phi\Big)\Bigg)=0.\right.\label{20}
\end{eqnarray}

The analysis can be simplified if we assume empty bulk models and
do not take into account contributions of any matter fields on the
brane ($\tau^{\mu}_{\mu}=0=\tau^{m}_{m}$). Besides, in braneworld
models in the BD theory, the cosmological constant is no longer
constant and generally it depends on the large extra dimension.
Therefore, let us take ($\Lambda=0$) for simplicity. Finally, note
that different choices of $\xi$ lead to different contributions,
so we begin with $\xi=-1$ since it eliminates the overall warp
factor on the left-hand side of equation (\ref{20}). With these
several simplifications, the eq. (\ref{20}) reads
\begin{eqnarray}
\oint
\Bigg(\frac{2w}{\phi^{2}}\nabla^{m}\phi\nabla_{m}\phi+\frac{2}{\phi}\nabla^{m}\nabla_{m}\phi-4\tilde{R}+W^{-2}\bar{R}
\Bigg)=-\frac{8\pi(57+10w)}{3+2w}\sum_{i}T_{4}^{(i)}\oint
\frac{\Delta^{(1)}(r-r_{i})}{\phi}.\label{21}
\end{eqnarray}

Note the appearance of the Euler character
$\chi=\frac{1}{4\pi}\oint \tilde{R}$. The internal space of the
model can be characterized by $\chi$ and, then, for each model
this topological invariant contributes in a specific way to the
sum rules. For the above case, we have explicitly
\begin{eqnarray}
\oint
\Bigg(\frac{w}{\phi^{2}}\nabla^{m}\phi\nabla_{m}\phi+\frac{1}{\phi}\nabla^{m}\nabla_{m}\phi+\frac{W^{-2}}{2}\bar{R}
\Bigg)=8\pi\chi-\frac{4\pi(57+10w)}{3+2w}\sum_{i}T_{4}^{(i)}L_{i}\phi^{-1}(r_{i}),\label{22}
\end{eqnarray}
where $L_{i}$ is the area of the $S^{1}$ cycle (the extra
dimension compactified on the brane) and $\phi(r_{i})$ is the
dilaton field value on the $i^{th}$-brane at the $r_{i}$ position.
It is time to make a brief comment about this result. In the ref.
\cite{AG2}, it was raised the hypothesis of an inconsistence
between braneworld models and ``pure'' BD gravity (where $w$ is
expected to be $\sim 1$). The reason was because of the
$(w-1)\lambda$ coupling - where $\lambda$ is the brane tension -
which appears in an ubiquitous way in the effective Einstein
equation projected on the brane. In the context of this work (as
well as in ref. \cite{AG2}), the BD gravity acts as an
intermediate theory between General Relativity and an effective
gravity recovered from the  low energy limit of the string theory,
so the $(w-1)\lambda$ coupling is not a real problem in this
scope. However, for a ``pure'' BD gravity it seems to be a
problem, since one cannot define the contribution of the brane
vacuum energy to the projected Einstein equation. As one can see,
from eq. (\ref{22}), it is just an apparent inconsistence, since
there is not a definite forbidden value to the BD parameter.

Another interesting choice for the $\xi$ parameter is $\xi=3$. If
we reconsider all the simplifications which lead to (\ref{21}) but
with $\xi=3$ it results \be\frac{3(3+2w)}{8\pi}\oint
W^{2}\bar{R}=(6w-47)\sum_{i}T_{4}^{(i)}W^{4}(r_{i})\phi^{-1}(r_{i})L_{i}.\label{23}\ee
Note that there is no contribution from the derivatives of the
scalar field, as well as, from the Euler number. We remark that,
for a negative  scalar curvature constant it is possible to have a
multiple brane scenario only with negative brane tensions! On the
other hand, for a positive scalar curvature constant, it possible,
at least, one negative brane tension. We shall comment the results
in the next Section.

\section{Concluding Remarks}

We generalize the braneworld sum rules to the BD gravity.  In the
interface between General Relativity and an effective gravity
obtained from the low energy limit of the string theory, such a
generalization is quite necessary. The results recover the
previous case when the limit $w\rightarrow \infty$ is taken. The
main goal of obtaining consistence conditions in braneworlds is
the large application to several models.

On one hand, the constraints imposed by the equation (\ref{18}),
for example, must be respected by any braneworld model in the BD
gravity in which the scalar field depends only on the large extra
dimension and should be taken into account in more involved
models. It is a strong imposition. On the other hand however, as
the eq. (\ref{18}) shows, such a constraint is not to much
restrictive itself. Of course, the consistence conditions are much
more attractive if one assumes a particular model.

It is important to remark the role of the scalar field in the sum
rules. It is clear that, except in those cases showed by eq.
(\ref{23}), the shape of the scalar field is strongly related with
the consistence of the model. In this context, it is surprising
that there is, at least, one solution to the dilaton field
(\ref{19}) which reduces this importance. Note that this result is
not obvious from the Einstein-Brans-Dicke equation of motion
(\ref{8}).

To the particular cases analyzed from eq. (\ref{21}), it is
important to stress  the possibility of existing only positive or
only negative brane tensions. It is in sharp contrast with the
Randall-Sundrum model \cite{RSI}, where the tension of the two
branes are necessarily equal and opposite. This type of fine
tuning between the brane tensions is not a necessity for other
models too (see ref. \cite{Leblond}). Besides, as it was shown,
the sum rules did not impose any restriction on the BD parameter,
solving the apparent inconsistence, raised in ref. \cite{AG2},
between braneworld models and ``pure'' BD gravity.

To end with, we call the attention to the possibility of the sum
rules bringing information about the modulus stabilization problem
as well as the stabilization of the scalar field. It can be
obtained, in principle, by the analysis of the right magnitude of
the last term of the right-hand side of eqs. (\ref{22}) and
(\ref{23}), for instance. Of course, it does not provide a
mechanism of stabilization but it can help to point  a concrete
research direction. We shall address to these questions in a
future work.

\section*{Acknowledgments}

J. M. Hoff da Silva thanks to CAPES-Brazil for financial support.
M. C. B. Abdalla and M. E. X. Guimar\~aes acknowledge CNPq for
partial support.

\end{document}